\documentclass[a4paper,11pt]{article}
\usepackage{pos}

\newcommand{\Fbg}{F_\mathrm{bg}}
\newcommand{\Vbreak}{V_\mathrm{break}}
\newcommand{\Vover}{V_\mathrm{over}}
\newcommand{\Vbias}{V_\mathrm{bias}}

\usepackage{hyperref}
\usepackage[nooneline]{subfigure}
\subfiguretopcaptrue

\title{Study on the gain and photon detection efficiency drops of silicon photomultipliers under bright background conditions}
\ShortTitle{SiPM gain drop under bright background conditions}

\author*[a,b,c]{Akira~Okumura}
\author[a]{Kawori~Wakazono}
\author[a]{Kazuhiro~Furuta}
\author[a,b,c]{Hiroyasu~Tajima}

\affiliation[a]{Institute for Space--Earth Environmental Research, Nagoya University,\\Furo-cho, Chikusa-ku, Nagoya 464-8601, Japan}
\affiliation[b]{Kobayashi--Maskawa Institute for the Origin of Particles and the Universe, Nagoya University,\\Furo-cho, Chikusa-ku, Nagoya 464-8602, Japan}
\affiliation[c]{Nagoya University Southern Observatories, Nagoya University,\\Furo-cho, Chikusa-ku, Nagoya 464-8602, Japan}

\emailAdd{oxon@mac.com}

\abstract{The use of silicon photomultipliers (SiPMs) in imaging atmospheric Cherenkov telescopes is expected to extend the observation times of very-high-energy gamma-ray sources, particularly within the highest energy domain of 50–300 TeV, where the Cherenkov signal from celestial gamma rays is adequate even under bright moonlight background conditions. Unlike conventional photomultiplier tubes, SiPMs do not exhibit quantum efficiency or gain degradation, which can be observed after long exposures to bright illumination. However, under bright conditions, the photon detection efficiency of a SiPM can be undergo temporary degradation because a fraction of its avalanche photodiode cells can saturate owing to photons from the night-sky background (NSB). In addition, the large current generated by the high NSB rate can increase the temperature of the silicon substrate, resulting in shifts in the SiPM breakdown voltages and consequent gain changes. Moreover, this large current changes the effective bias voltage because it causes a voltage drop across the protection resistor of 100--1000\,$\Omega$. Hence, these three factors, namely the avalanche photodiode (APD) saturation, Si temperature, and voltage drop must be carefully compensated for and/or considered in the energy calibration of Cherenkov telescopes with SiPM cameras. In this study, we measured the signal output charge of a SiPM and its variation as a function of different NSB-like background conditions up to 1 GHz/pixel. The results verify that the product of the SiPM gain and photon detection efficiency is well characterized by these three factors.}

\ConferenceLogo{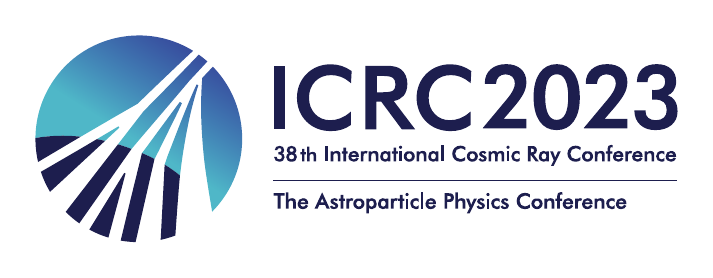}

\FullConference{%
The 38th International Cosmic Ray Conference (ICRC2023)\\
  26 July -- 3 August, 2023\\
  Nagoya, Japan}

\begin{document}
\maketitle

\section{Introduction}

The characterization of photodetectors used in gamma-ray and cosmic-ray telescopes is a crucial calibration item in the entire optical system. To accurately reconstruct gamma-ray and cosmic-ray energies from the number of detected Cherenkov or fluorescence photons, the photodetector gains and photon detection efficiencies (PDEs) must be well calibrated for different supply voltages and photon wavelengths.

Typically, these calibrations are first performed under a dark condition in the laboratory, following which the characteristics are assessed again under different background illumination levels within the same system or using the telescopes. This is because the gamma-ray or cosmic-ray sky is not free of night-sky background (NSB) photons emitted by stars, airglow, or the Moon. Therefore, possible changes in gain and PDEs must be monitored under different brightness conditions.

In addition to conventional photomultiplier tubes (PMTs), the use of silicon photomultipliers (SiPMs) in gamma-ray and cosmic-ray telescopes is being studied. For example, the Small-Sized Telescopes (SSTs) of the Cherenkov Telescope Array (CTA) will employ 2048 SiPM pixels (6\,mm $\times$ 6\,mm each) to form their ${\sim}320$\,mm spherical focal plane with a field of view of about $9^\circ$ \cite{Depaoli:2023:Status-of-the-SST-Camera-for-the-Cherenkov-Telesco}.

The main physics objective of the CTA SSTs is to discover and observe galactic objects called PeVatrons that accelerate cosmic rays up to the ``knee'' energy. The SSTs utilize the imaging atmospheric Cherenkov telescope (IACT) technique with a 4\,m Schwarzschild--Couder optical system and a SiPM camera to detect celestial gamma rays with energies ranging from 5~TeV to beyond 300\,TeV \cite{Cherenkov-Telescope-Array-Consortium:2018:Science-with-the-Cherenkov-Telescope-Array}. To maximize the discovery potential of SSTs, particularly in the highest energy range above a few tens of TeV for the PeVatron search, long observation times on the order of 100 hours are desirable. Accordingly, the highest-energy observations will be performed even under bright moon conditions, which do not severely affect the SST performance at the highest energies.

The NSB count rate of a single SST camera pixel is expected to be about 40\,MHz under moonless, dark night conditions. However, this value can increase up to about 1\,GHz under full moon conditions, leading to possible systematic uncertainties in the triggering and energy calibration of SSTs. Thus, we need to understand the behavior of the SiPMs operated at a constant high voltage under such conditions to make the SST energy calibration more accurate and eliminate possible energy biases that can occur in different background brightness.

Here, we first discuss three possible causes of SiPM gain and PDE changes in Section~\ref{sec:reasons} to list the characteristics of SiPMs or individual avalanche photodiode (APD) cells that need to be investigated. In Sections~\ref{sec:brightness_temperature} and \ref{sec:drops}, we present the results of our SiPM measurements under different brightness conditions, verifying that the aforementioned causes can explain the SiPM behavior.

\section{Possible Causes of SiPM Gain and PDE Drops}
\label{sec:reasons}

Three possible factors can decrease the gain and PDE of a SiPM under bright NSB conditions. First, the bias voltage applied to a SiPM effectively drops owing to the large current and voltage drop across a series protection resistor. Assuming an NSB rate of 1\,GHz, a protection resistor of 1\,k$\Omega$, and a gain of $5\times10^6$, the current and voltage drop would be 0.8\,mA and 0.8\,V, respectively, resulting in a more than 10\% gain drop for a SiPM operating at an overvoltage ($\equiv$ bias voltage $-$ breakdown voltage) of 6\,V. The gain of a SiPM is approximately proportional to overvoltage; hence, the gain drop is calculated to be $0.8/6 \sim 0.13$\%.

The second causative factor is the temperature increase of the SiPM, which can not be adequately cooled when the NSB rate is high. With the assumptions in the previous paragraph and a breakdown voltage of around 50\,V, the heat generated in a single SiPM pixel reaches ${\sim}$0.04\,W, which is sufficient to decrease the SiPM gain. This is because the SiPM breakdown voltage is a function of the temperature, and its dependence is $0.033\pm0.001$\,V/$^\circ$C for the product used in the current study (Hamamatsu Photonics S14520-6050-VS, $6\times6$\,mm$^2$, 50\,$\mu$m APD cell size). Therefore, a temperature increase of 3\,$^\circ$C can decrease the gain by $0.1/6\sim0.17$\%.

In addition to the two aforementioned factors, the gain of individual APD cells may decrease when the NSB rate is high, and the number of APD cells is not sufficiently large. As the APD cells behave as parallel capacitors in a single SiPM pixel, the photoelectron pulse signals of the SiPM can be considered as their discharge signals with a time constant on the order of 0.1--1\,$\mu$s (recovery time). Therefore, a high NSB rate can result in a non-negligible fraction of APD cells that have already generated avalanche multiplications or are in the process of discharging. If Cherenkov photons enter these ``recovering'' cells, the effective PDE and gain may be lower than those of fully charged cells. Assuming 14,400 APD cells (50\,$\mu$m cells in a 6\,mm pixel), 1\,GHz NSB, and a recovery time of 0.1\,$\mu$s, roughly 1\% of the APD cells would be in the recovery state.

\section{Background Brightness and SiPM Temperature Measurements}
\label{sec:brightness_temperature}

\begin{figure}
  \centering
  \makebox[\textwidth][c]{
  \subfigure[]{%
    \includegraphics[width=.55\textwidth,clip]{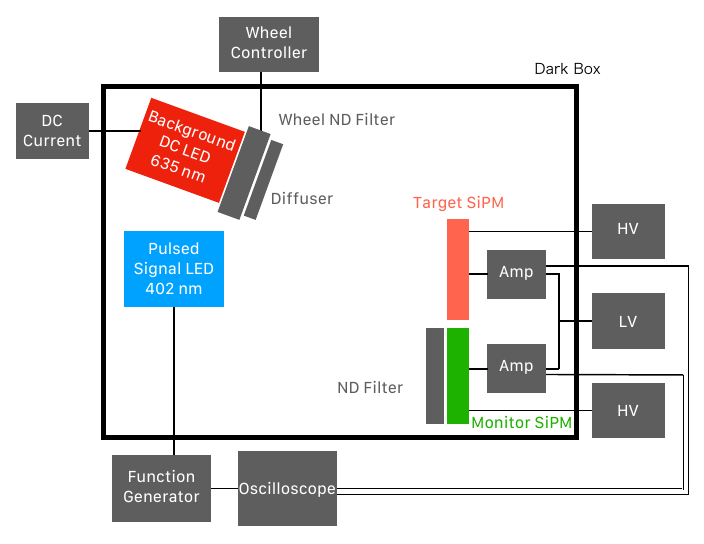}
    \label{fig:darkbox2}
  }%
  \subfigure[]{%
    \includegraphics[width=.45\textwidth,clip]{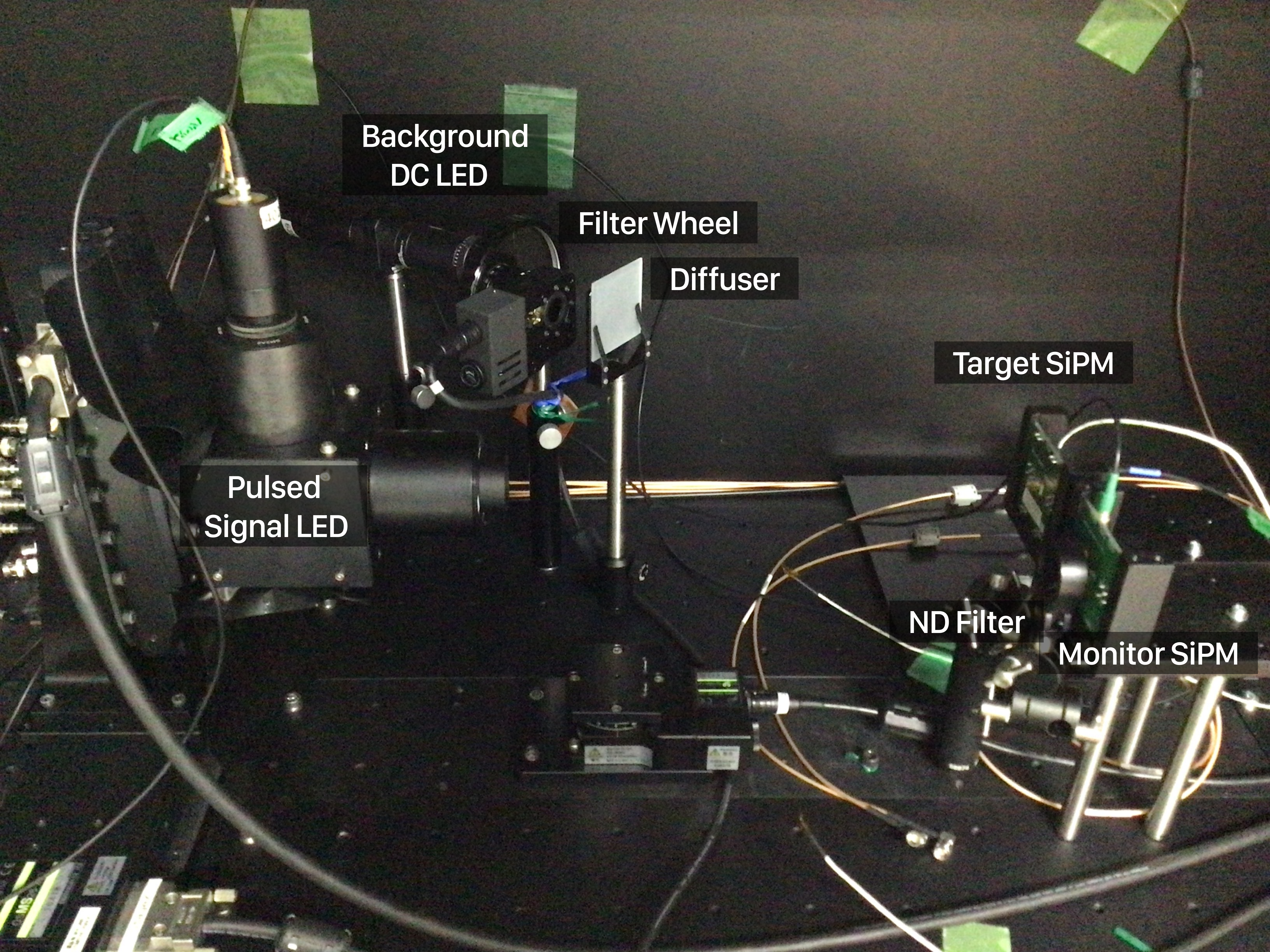}
    \label{fig:darkbox1}
  }
  }
  \caption{(a) Schematic view of the measurement system. The target and monitor SiPMs are connected to amplifiers, and the output is fed to an oscilloscope synchronized with a function generator. (b) Photograph of the system.}
\end{figure}

To verity the rough estimates in the previous section and to specify future SST calibration items, we integrated the output signal voltage of a SiPM with a controlled amount of pulsed light emitted under different background conditions. A single-pixel SiPM (referred to as ``target SiPM''), S14520-6050-VS, connected to series resistors with a total resistance of 57\,$\Omega$, a pulsed light source (402\,nm LED), and a DC light source (635\,nm LED) were placed in a dark box at a room temperature ($25.2\pm0.1^\circ$C) as illustrated in Figs.~\ref{fig:darkbox2} and \ref{fig:darkbox1}. By varying the brightness of the DC light, i.e., the count rate of the target SiPM, from 0\,GHz to ${\sim}1$\,GHz, which was monitored using another SiPM (``monitor SiPM,'' Hamamatsu Photonics S13360-6050-CS), we derived the relative gain change of the target SiPM as a function of the background level.

\begin{figure}
  \centering
  \makebox[\textwidth][c]{
  \subfigure[]{%
    \includegraphics[width=.5\textwidth,clip]{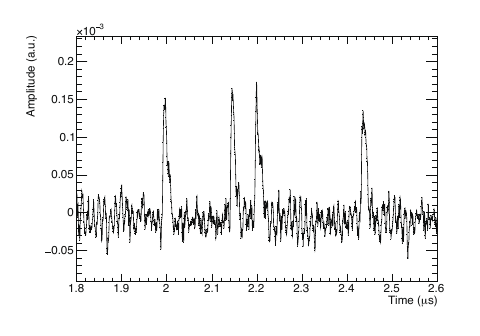}
    \label{fig:oscilloscope}
  }%
  \subfigure[]{%
    \includegraphics[width=.5\textwidth,clip]{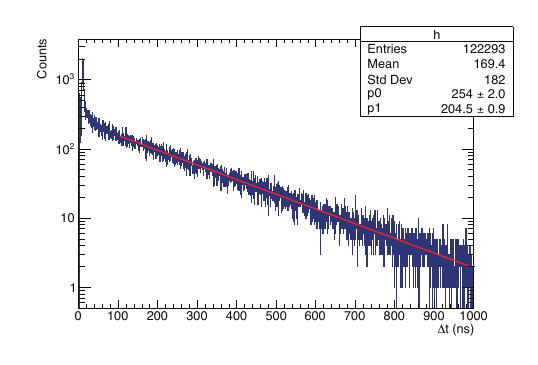}
    \label{fig:deltaT}
  }
  }
  \caption{(a) SiPM waveform example after an offline waveform shaping was performed. The four peaks are single-photoelectron pulses generated by the background photons. (b) $\Delta t$ distribution of single-photoelectron events (histogram), where an exponential distribution appears in the large $\Delta t$ region, which can be fitted with an exponential function (red line).}
\end{figure}

The background brightness, which is the most basic parameter in this measurement, was first calibrated by measuring the distributions of the time differences between two photoelectron counts ($\Delta t$). As long as the count rate is low enough to resolve two consecutive pulses, the SiPM output recorded by the oscilloscope looks like the example in Fig.~\ref{fig:oscilloscope}. By fitting the exponential tail of the $\Delta t$ distribution, the background count rate can be calculated. However, at the brightest setting of 1\,GHz, single-photoelectron pulses could not be separated from the target SiPM output. Instead, the output of the monitor SiPM with a pre-calibrated ND filter was analyzed to estimate the background level of the target SiPM.

The temperatures of the SiPM under different brightness levels are difficult to measure directly. As the target SiPM has an active area of only $6\times6$\,mm$^2$, it is impossible to place a thermometer on the SiPM surface without blocking the incident photons. Therefore, we employed an indirect method to estimate the SiPM temperature.

The dark current of the target SiPM can be measured as a function of the ambient temperature when the dark count rate is low, and the temperature difference between the SiPM and air is small. The dark current vs. temperature of the target SiPM was first calibrated in a dark thermal chamber to estimate the SiPM temperature in the dark box. During the background count rate measurements, the DC background LED illuminating the target SiPM was turned off at $t=t_0$, and the temporal change in the dark current was measured using the sourcemeter (Keithley 2400) used as the bias voltage supply.

\begin{figure}
  \centering
  \makebox[\textwidth][c]{
  \subfigure[]{%
    \includegraphics[width=.5\textwidth,clip]{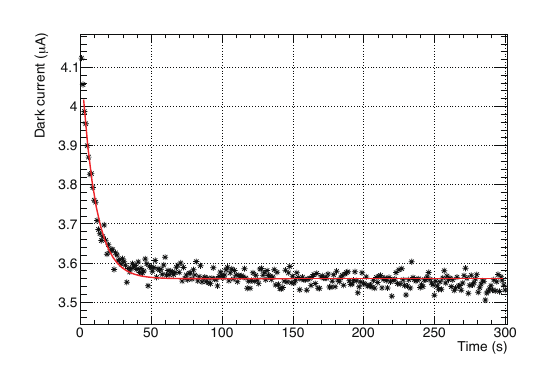}
    \label{fig:current1}
  }%
  \subfigure[]{%
    \includegraphics[width=.5\textwidth,clip]{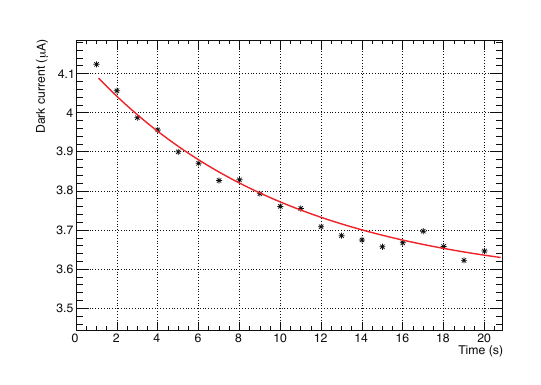}
    \label{fig:current2}
  }
  }
  \caption{(a) Example of the temporal change in the dark current of the target SiPM. An exponential fit is depicted with a red curve. (b) Same as (a) but the first 20\,s are zoomed in.}
\end{figure}

As illustrated in Figs.~\ref{fig:current1} and \ref{fig:current2}, the dark current decreases as an exponential function, because the SiPM temperature approaches the dark box ambient temperature after the background illumination is turned off. The dark current at $t_0$ estimated by the exponential fit can be considered as the current while the LED is on; hence, the SiPM temperature can also be estimated by using the dark current vs. temperature function calibrated beforehand. In addition, the breakdown voltage at each background level can also be calculated by separately measuring the breakdown voltage at different ambient temperatures in a thermal chamber.

\section{Measurement of SiPM Gain and PDE Drops}
\label{sec:drops}

Using the SiPM current, $I$, measured by the sourcemeter; the background rate, $\Fbg$; and the shift in the breakdown voltage due to temperature rise, $\Delta \Vbreak$, the effective overvoltage, $\Vover$, can be expressed as a function of $\Fbg$.
\begin{equation}
  \Vover(\Fbg) = \Vbias - I(\Fbg) \times 57\,\Omega - \Vbreak (T(\Fbg)),
\end{equation}
where $\Vbias$ denotes the fixed bias voltage of 42.8\,V through this study, $\Vbreak (T=25^\circ\mathrm{C}) = 39.0$\,V represents the pre-calibrated breakdown voltage at $T=25^\circ\mathrm{C}$.

\begin{figure}
  \centering
  \includegraphics[width=.75\textwidth,clip]{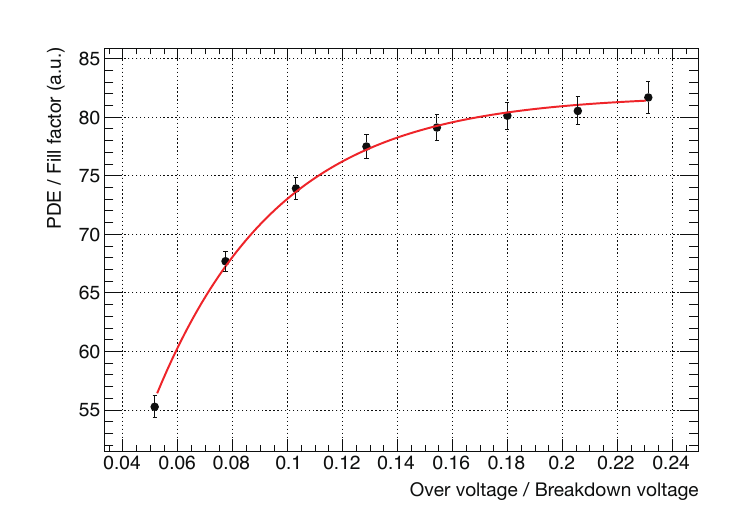}
  \caption{Relative PDE change in the target SiPM as a function of $\Vover/\Vbreak$.}
  \label{fig:PDE}
\end{figure}

Considering that the PDE, $\epsilon$, is dependent on $\Vover / \Vbreak$  with a function $\epsilon \propto 1-\exp(-C\Vover/\Vbreak)$ \cite{Otte:2017:Characterization-of-three-high-efficiency-and-blue}, as illustrated in Fig.~\ref{fig:PDE}, the relative charge of the SiPM output for a fixed amount of pulsed LED flash can be expressed as
\begin{equation}
  \frac{\bar{Q}(\Fbg)}{\bar{Q}(\Fbg=0)} = \frac{\Vover(\Fbg) \left \{1 - \exp(-C\Vover(\Fbg)/\Vbreak(T(\Fbg))) \right \}}{ \Vover(\Fbg = 0) \left \{ 1 - \exp(-C\Vover(\Fbg=0)/\Vbreak| _{T=25^\circ\mathrm{C}}) \right \} },
\end{equation}
where $C$ represents a fit parameter.

\begin{figure}
  \centering
  \includegraphics[width=.75\textwidth,clip]{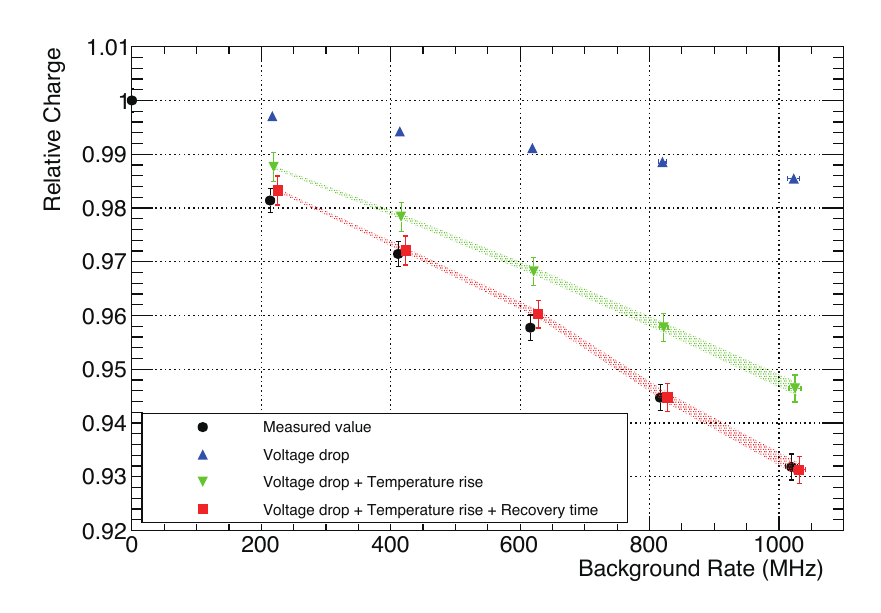}
  \caption{Relative drop in the integrated charge of the target SiPM output at different background rates. Measured data (black circles) are depicted with the expected drops from the three causes. The error bars represent statistical uncertainties, and the shaded bands depict the systematic uncertainty in $\Vbreak(T)$. Green and red data points are intentionally shifted to avoid overlaps.}
  \label{fig:drop}
\end{figure}

By integrating the oscilloscope waveform in the LED flash time window, its average $\bar{Q}(\Fbg)$ can be compared with the $\bar{Q}(\Fbg=0)$ measured with the background LED off. Fig.~\ref{fig:drop} demonstrates how the relative charge $\bar{Q}(\Fbg)/\bar{Q}(\Fbg=0)$ decreases as the background rate $\Fbg$ increases.

The effects of the voltage drop across the series resistors and the temperature increase are also compared in Fig.~\ref{fig:drop}. However, the observed relative charge drop cannot be fully explained by these factors alone; hence, we must also consider the recovery time as discussed in Section~\ref{sec:reasons}.

The effective overvoltage of a single APD cell drops and recovers with the following exponential function;
\begin{equation}
  V_\mathrm{over, recovery}(t) = \Vover \left \{ 1 - \exp(-t/0.095\,\mu\mathrm{s} )\right \},
\end{equation}
where the time constant of 0.095\,$\mu$s is obtained by fitting the pulse output of the target SiPM. The effective gain of a single APD cell is proportional to the effective $\Vover$.

Using the above time profile, number of APD cells, and background rate, we performed a toy Monte Carlo simulation to compute the product of the average relative gain and PDE. Considering this last factor, the relative charge drop presented in Fig.~\ref{fig:drop} is finally explained consistently within the observed statistical errors.

\section{Discussion and Conclusion}

In this study, we have measured the relative output charge of a SiPM under different brightness conditions ranging from 0 to 1\,GHz. The results verify that the charge is proportional to the SiPM gain and PDE and that the drop can be well explained by three factors: the voltage drop across the series protection resistor(s), temperature increase of the SiPM, and non-negligible recovery time of the individual APD cells.

In these measurements, a rather small protection resistor with a resistance of 57\,$\Omega$ was intentionally employed to clearly separate the three causes. Therefore, the contribution of the voltage drop across the resistor is not dominant in Fig.~\ref{fig:drop}. However, the final SST camera design will use a 1\,k$\Omega$ resistor, which will increase this contribution by a factor of ${\sim}20$. Therefore, it is important to accurately monitor the current of each SiPM pixel to calculate the effective bias voltage during observations.

Compared to the first factor, the other two factors, i.e., the temperature and recovery, are less dominant in the case of SSTs but still not negligible. To account for the effects of these factors, we need to calibrate and understand the camera pixel behavior at different background levels in the laboratory prior to gamma-ray observations.

\providecommand{\href}[2]{#2}\begingroup\raggedright\endgroup

\acknowledgments

This study was supported by JSPS KAKENHI Grant Numbers JP18KK0384, JP20H01916, and JP23H04897.

%
%
%

\end{document}